\documentclass{webofc}
\usepackage[varg]{txfonts}

\newcommand{\Xp}{\Xi_{cc}^+}
\newcommand{\Xpp}{\Xi_{cc}^{++}}
\newcommand{\Lc}{\Lambda_c^+}
\newcommand{\Jp}{J/\psi}

\newcommand{\MeV}{\mbox{\,MeV}}
\newcommand{\GeV}{\mbox{\,GeV}}
\newcommand{\TeV}{\mbox{\,TeV}}

\newcommand{\fb}{\mbox{\,fb}}
\newcommand{\fs}{\mbox{\,fs}}
\newcommand{\fm}{\mbox{\,fm}}

\begin{document}
\title{Resolving the SELEX–LHCb double-charm baryon conflict:
the impact of intrinsic heavy-quark hadroproduction and supersymmetric
light-front holographic QCD}

\author{\firstname{Sergey} \lastname{Koshkarev}\inst{1}\fnsep%
\thanks{\email{sergey.koshkarev@ut.ee}}
\and
\firstname{Stefan} \lastname{Groote}\inst{1}\fnsep\thanks{\email{groote@ut.ee}}
}

\institute{Institute of Physics, University of Tartu, 51010, Estonia}

\abstract{%
In this paper we review the hadroproduction mechanisms of double-charm
baryons for the different experimental environments and reinterpret the SELEX
and LHCb results.
}
\maketitle
\section{Introduction}
The SELEX measurements of the production of the double-charm baryons at
large $x_F$ $(\langle x_F\rangle \sim 0.33)$ are among the most intriguing and
surprising results in modern baryonic physics~\cite{Mattson:2002vu,%
Mattson:2002dz,Ocherashvili:2004hi}. The SELEX experiment was a fixed-target
experiment utilizing the Fermilab negative charged beam at 600~GeV$/c$ to
produce charm particles in a set of thin foil of Cu or in a diamond and was
operated in the kinematic region $x_F > 0.1$. The negative beam composition
was about 80\% $\Sigma^{-}$ and 20\% $\pi^{-}$. The positive beam contained
90\% protons. In a first observation using the sample of
$\Lc\to pK^-\pi^+$~\cite{Kushnirenko:2000ed,Kushnirenko}, SELEX found a signal
of $15.9$ events over $6.1\pm 0.1$ background events in the channel 
$\Xp\to\Lc K^-\pi^+$~\cite{Mattson:2002vu}. To complement this result, SELEX
published an observation of $5.62$ signal events over $1.38\pm 0.13$
background events for the decay mode $\Xp\to pDK^-$ from a sample of
$D^+\to K^-\pi^+\pi^+$ decays~\cite{Ocherashvili:2004hi}.

Two charm quarks at high $x_F$ cannot be produced from DGLAP
evolution~\cite{Gribov:1972ri,Dokshitzer:1977sg,Altarelli:1977zs} or
perturbative gluon splitting
$g\to g+g\to(\bar cc)+(\bar cc)$~\cite{Altarelli:1977zs,Curci:1980uw,%
Furmanski:1980cm}. Therefore, the observation of a double-charm baryon
$|qcc\rangle$ at a large mean value for $x_F$ and a relatively small
mean transverse momentum by SELEX can raise skepticism. However,
$\Lambda_c(udc)$ and $\Lambda_b(udb)$ were both discovered at the ISR at high
$x_F$~\cite{Lockman:1979aj,Chauvat:1987kb,Bari:1991ty}. In addition, the NA3
experiment measured both the single-quarkonium hadroproduction
$\pi A\to J/\psi X$~\cite{Badier:1983dg} and the double-quarkonium
hadroproduction $\pi A\to J/\psi J/\psi X$~\cite{Badier:1982ae} at high $x_F$.
In fact, all of the events $\pi A\to J/\psi J/\psi X$ were observed by NA3
with a total value of $x_F>0.4$. 

Recently, the LHCb collaboration published an observation of $313\pm 33$
events of $\Xpp\to\Lc K^-\pi^+\pi^+$ in a $13\TeV$ sample at the LHC and
$113\pm 21$ events in a $8\TeV$ sample at mass $3621.40\pm 0.72\text{(stat)}
\pm 0.27\text{(sys)}\pm 0.14(\Lc)\MeV/c^2$, corresponding to $1.7\fb^{-1}$ and
$2\fb^{-1}$, respectively~\cite{Aaij:2017ueg}. This result was again
complemented by the mode $\Xpp \to \Xi_{c}^{+}\pi^{+}$~\cite{Aaij:2018gfl}.
LHCb reported that the mass difference between the $\Xp(dcc)$ candidate
reported by SELEX and the $\Xpp(ucc)$ state reported by LHCb was
$103\MeV/c^2$. Therefore, these states cannot be readily interpreted as an
isospin doublet since one would expect a mass difference of isospin partners
of only a few $\MeV/c^2$. Note, though, that the upper limit of the $x_F$
range at the LHCb collider experiment is given by $x_F\approx 0.15$ and
$x_F\approx 0.09$ for the $8\TeV$ and $13\TeV$ analysis, respectively. In
contrast to this, the $x_F$ range at the SELEX fixed-target experiment starts
at $x_F=0.1$, nearly complementary to the acceptance for the LHCb. The
lifetime of $\Xpp$  state measured by LHCb is
$256^{+24}_{-22} \pm 14\fs$~\cite{Aaij:2018wzf}. The upper limit of the
lifetime measured by SELEX is given by $\tau(\Xp) < 33\fs$. Again, these
results are in contradiction, $\tau(\Xpp) / \tau(\Xp) \approx 2.5 - 4$.

\section{Production rate and the kinematics of the $\Xp$\\
  for the SELEX experiment}

The SELEX collaboration did not provide the absolute production rate for the
double charmed baryon state $|dcc\rangle$. Fortunately, this rate can be
compared to that of $\Lc$ baryon. The production ratio $R_{\Lc}$ measured by
SELEX is given by
\[
R_{\Lc}^{\text{SELEX}} = \frac{\sigma(\Xp) \cdot
  Br(\Xp \to \Lc K^- \pi^+) }{\sigma(\Lc)}
  =\frac{N_{\Xp}}{\epsilon_+} \cdot \frac{\epsilon_{\Lc}}{N_{\Lc}},
\]
where $N$ is the number of events in the respective sample, and the
reconstruction efficiency of $\Xp$ is given by
$\epsilon_+\approx 11\%$~\cite{Mattson:2002vu}. The central value for the
number $N_{\Lc}/\epsilon_{\Lc}$ of reconstructed $\Lc$ baryon events reported
in Ref.~\cite{Garcia:2001xj} lies between $13326$ and $10010$ according to
whether the lowest bin with $x_F\in [0.125,0.175]$ is taken into account or
not.  If we take into account the intrinsic charm mechanism, the reconstruction
efficiency of $\Xp$ will grow at least by a factor of $2.3$ mainly because the
$x_F$ distribution predicted by the intrinsic charm mechanism at large Feynman
$x_F$ is well matched to the acceptance of the SELEX fixed-target
experiment~\cite{Mattson:2002dz} (cf.\ Ref.~\cite{Groote:2017szb}).
Therefore, we obtain
\[
R_{\Lc}^{\rm SELEX}\sim(0.5-0.6)\times 10^{-3}
\]
It is clearly of interest to relate the production of the $\Xp$ at the SELEX
experiment to the production of the double $J/\psi$ production at the NA3
experiment. Unfortunately, it is not possible to compare the two results
directly. However, we are able to compare the following ratios
$R=\sigma(c\bar cc\bar c)/\sigma (c\bar c)$:
\[
R^{\rm SELEX}=R_{\Lc}\times\frac{f(c\to\Lc)}{f_{\Xi_{cc}}}
  \sim(1-4)\times 10^{-3}
\]
\vspace{-7pt}\noindent
and
\vspace{-7pt}
\[
R^{\rm NA3}=\frac{\sigma(\psi\psi)}{\sigma(\psi)}\times
  \frac{f_\psi}{f^2_{\psi/\pi}}\sim 2\times 10^{-2},
\]
where $f_{\psi/\pi}\approx 0.03$ is the fragmentation rate of the intrinsic
charm state of the pion into $J/\psi$~\cite{Vogt:1995tf} and
$f_\psi\approx 0.06$ is the perturbative QCD fragmentation rate into
$J/\psi$~\cite{Mangano:2004wq}.
$f_{\Xi_{cc}}\approx 0.25$~\cite{Koshkarev:2016rci} represents the fraction of
double $c\bar c$ pairs producing the sum of single-charged baryons $\Xp$ and
double-charged baryons $\Xpp$, but this fraction cannot be less than the
fraction to produce $J/\psi$. Therefore, $R^{\rm SELEX}$ should not be larger
than $10^{-2}$. The SELEX production ratio is thus in approximate consistency
with the complementary measurement of the double $J/\psi$ production by the
NA3 experiment. It is interesting to note that the intrinsic charm mechanism
predicts $\langle x_F(\Xi_{cc})\rangle=0.33$, as shown in
Ref.~\cite{Koshkarev:2016rci}. This is in excellent agreement with the value
$\langle x_F(\Xp)\rangle\sim 0.33$ measured by the SELEX experiment.

\section{Mass difference}
In order to resolve the discrepancy between the results from SELEX and LHCb
we will utilize the predictions of the supersymmetric light front holographic
QCD (SUSY LFHQCD). This approach was developed by imposing the constraints
from the superconformal algebraic structure on LFHQCD for massless
quarks~\cite{Dosch:2015nwa}. As has been shown in
Refs.~\cite{Dosch:2015nwa,Dosch:2015bca}, supersymmetry holds to a good
approximation, even if conformal symmetry is strongly broken by the heavy
quark mass.

Note that the $3_C+\bar 3_C$ diquark structure of the $\Xp$ can be written
explicitly as state $|[dc]c\rangle$. The production of the double-charm baryon
state $\Xi_{[dc]c}^+$ with $[dc]$ in a spin-singlet state is natural in the
SELEX fixed target experiment since it has acceptance at high $x_F$, i.e.,\ in
the realm of intrinsic charm; the $[dc]c$ configuration can easily re-coalesce
from a higher Fock state of the proton such as $|uudc\bar cc\bar c\rangle$. In
contrast, the production of this state is likely to be suppressed in
$q\bar q\to c\bar cc\bar c$ or $gg\to c\bar cc\bar c$ reactions at the LHCb.
Thus LHCb has most likely observed the double-charm baryon state
$|u(cc)\rangle$.\footnote{We use square brackets $[\ ]$ for spin-0 and
round brackets $(\ )$ for spin-1 internal states.} The mass difference between
the $|[dc]c\rangle$ and the $|u(cc)\rangle$ states is due to the hyperfine
interaction between the quarks.

Supersymmetric light front holographic QCD, if extended to the case of two
heavy quarks, predicts that the mass of the spin-1/2 baryon should be the same
as the mass of $h_c(1P)(3525)$ meson~\cite{Dosch:2015bca}. This is well
compatible with the SELEX measurement of $3520.2\pm 0.7\MeV/c^2$ for the
$\Xp(d[cc])$, although the uncertainty of SUSY LFHQCD predictions is at least
of the order of $100\MeV$. Indeed, the mass of the $|u(cc)\rangle$ state is
predicted to be the same as that of the $\chi_{c2}(1P)(3556)$ meson, which is
in turn lower than the LHCb result of
$3621.40\pm 0.72\text{(stat)}\pm 0.27\text{(sys)}\pm 0.14(\Lc)\MeV/c^2$
for the $\Xpp$.

Supersymmetric LFHQCD is based on and best tested in the chiral limit of QCD,
where all quarks are massless. The mass difference between the $h_c(1P)$ and
the $\chi_{c2}(1P)$ is mainly due to the hyperfine splitting between the two
charm quarks, and hence very small. However, for a baryon there is a spin-spin
interaction between the charm quark and a light quark which is
larger than the hyperfine splitting.  By comparing hadron masses with the
masses of light and charmed quarks, one can estimate the strength of this
additional supersymmetry-breaking interaction to be found in the range
$84-136\MeV/c^2$~\cite{Dosch:2015bca, Brodsky:2016yod}, which is well
compatible with the mass difference between the SELEX and the LHCb states.

\section{The $\Xpp(3780)$ state}
At a few conferences~\cite{Moinester:2002uw,Engelfried:2007at} (cf.\ also the
PhD thesis of Mark E.~Mattson~\cite{Mattson:2002dz}), the SELEX collaboration
presented a decay process $\Xpp(3780) \to \Lc K^- \pi^+ \pi^+$ for the state
$\Xpp(3780)$ with statistical significance of $6.3\,\sigma$. By removing the
slower part of the $\pi^+$'s, SELEX observed that roughly $50\%$ of the signal
events above background decay weakly and $50\%$ decay strongly (to $\pi^+\Xp$).
However, this is not possible for a single state. As SELEX did not find a
plausible explanation for its decay properties, the result was not published.

Assuming $\Xpp(3780)$ to be an excited state $|ucc^*\rangle$, we predict the
mass of $|ucc^*\rangle$ by utilizing the predictions of supersymmetric light
front holographic QCD (SUSY LFHQCD). The SUSY LFHQCD prediction for the baryon
mass spectra is given by the simple formula
\begin{equation*}
M^2 \propto \lambda (n + L + 1) \,
\end{equation*}
where $\sqrt{\lambda}\approx 0.52\GeV$ is the fundamental mass parameter,
given by the characteristic mass scale of QCD~\cite{Brodsky:2016yod}. Using
this simple formula, we can estimate the masses of the states
$|[qc]c\rangle_{3/2}$ ($n=1$, $L=0$) and $|(qc)c\rangle_{3/2}$ ($n=0$, $L=1$),
where $(qc)$ indicates the spin-1 diquark. These states should have the same
mass around $3730\MeV/c^2$, where the uncertainty of the SUSY LFHQCD
predictions is at least of the order of $100\MeV$. Obviously, we have good
agreement with the data for $\Xpp(3780)$.

Investigating the decay properties, the $|[qc]c\rangle_{3/2}$ is more
preferable for the weak decay. In contrast to that, $|(qc)c\rangle_{3/2}$
includes a $D^*$-meson-like state, leading to the strong decay
$(qc)\to[qc]+\pi$, similar to $D^* \to D + \pi$.

\section{Isopin splitting of the SELEX states}
The analysis of the isospin splitting of the SELEX states implies that
double-charm baryons are very compact, i.e.\ the light quark must be very
close to the two heavy quarks~\cite{Brodsky:2011zs}. This contradicts the
usual wisdom. Indeed, within the heavy-diquark concept, the production of the
double-charm baryon can proceed in two steps. In a first step, due to the
reactions $q\bar q\to c\bar cc\bar c$ or $gg \to c\bar cc\bar c$ the
production of two charm quarks with a small relative momentum will take place,
followed by the formation of a $cc$-diquark in the color-antitriplet state. In
a second step, the transition of the produced diquark into the baryon is
performed. The normalization of the fragmentation of the $cc$-diquark into the
double-charm baryons is unknown. However, one is still able to provide some
quantitative analysis because the fragmentation function is proportional to
the wave function at the origin. The color-antitriplet wave function can be
estimated on the basis of information about the color-singlet wave function,
$|R(0)[cc]_{\bf\bar 3}|\sim|R(0)[c\bar c]_{\bf 1}|$. This leads to an
atom-like structure where the $cc$ diquark forms the compact core while the
scale of the light quark is given by the nonperturbative confinement
scale~\cite{Kiselev:2001fw}. In case of the $S$-wave solution we have the
scale hierarchy
\begin{equation*}
r_{cc} : r_{QCD} \approx 0.39 : 1 \,
\end{equation*}
where $r_{cc}\sim r_{\Jp}\sim 0.39\fm\simeq(0.5\GeV)^{-1}$~\cite{Nochi:2016wqg}
and $r_{QCD}=1/(\Lambda_{QCD}\approx 200\MeV) \approx 1\fm$. 

In contrast to this, the production of the double-charm baryons at the SELEX 
experiment is supposed to be due to re-coalesce from a higher Fock state of
the proton such as~\cite{Koshkarev:2016rci,Brodsky:2017ntu}
\begin{equation*}
|[uu]_{\bf\bar 3}[dc]_{\bf\bar 3}c_{\bf 3}[\bar c \bar c]_{\bf 3}\rangle
\end{equation*}
which leads to the baryonic state $|[qc]c\rangle$ in a natural way. Here the
scale will be characterized by the size of the spin-0 $[qc]$ diquark, given
by the Compton wavelength $\lambda_{[qc]}\sim 1/m_{[qc]}$ of the diquark.
This naturally provides closeness of the light quark to the two heavy quarks.
Note that the peculiarities mentioned here are due to the inclusion of two
heavy quarks.

It is interesting to estimate the compactness of such state. The mass
$m_{[qc]}$ can be estimated as the effective diquark mass,
$m_{\Xi_{cc}}-m_{c}$, where $m_{\Xi_{cc}}$ is the double-charm baryon mass
and $m_{c}$ is the mass of the $c$ quark. In case of the SELEX $3520\MeV$
event, one has $\lambda_{[dc]} \sim 0.5\fm$ which is again in the perfect
agreement with the compactness of the SELEX state calculated from isospin
splitting~\cite{Brodsky:2011zs}.

\section{Summary}
Using both theoretical and experimental arguments, we have shown that the
SELEX and the LHCb results for the production of double-charm baryons can
both be correct.  We have compared the data for double $J/\psi$ production
observed by the NA3 experiment and the SELEX result for $\Xp$ production at
high Feynman-$x_F$. We have found that the NA3 data strongly complement the
SELEX production rate for the spin-1/2 $|[dc]c\rangle$ state. In contrast,
LHCb has most likely discovered the the heavier state $|u(cc)\rangle$
produced by gluon--gluon fusion $gg\to c\bar cc \bar c$ at $x_F\sim 0$. The
application of supersymmetric algebra to hadron spectroscopy, together with
the intrinsic heavy-quark QCD mechanism for the hadroproduction of heavy
hadrons at large $x_F$, can thus resolve the apparent conflict between
measurements of double-charm baryons by the SELEX fixed-target experiment and
the LHCb experiment at the LHC collider. The mass difference of the two
double-charm baryons reflects the distinct spins of the underlying diquarks.

The theoretical estimate for the compactness of the baryon is in good
agreement with similar estimates from isospin splitting. In addition, we gave
estimates for the mass and the decay properties of
$\Xpp(3780) \to \Lc K^- \pi^+ \pi^+$~\cite{Koshkarev:2018kre}.
The production of the double-charm baryons at AFTER@LHC with 
the intrinsic charm is investigated in Ref.~\cite{Koshkarev:2016acq}.

There is still an unexplained difference in lifetimes. However, it is
interesting to note that the LHCb result
$\tau(\Omega_c^0) = 268 \pm 24 \pm 10 \pm 2\fs$~\cite{Aaij:2018dso} 
for the lifetime is again in disagreement with the respective lifetime 
measured by fixed-target experiments.
\begin{center}\begin{tabular}{ |c|c|c| } 
 \hline
 Experiment & lifetime (fs) & Number of events \\
  \hline
 FOCUS~\cite{Link:2003nq} & $72 \pm 11 \pm 11$ & 64 \\ 
 WA89~\cite{Adamovich:1995pf} & $55 ^{+13 \, +18}_{-11 \, -23} $  & 86\\ 
 E687~\cite{Frabetti:1995bi} & $86^{+27}_{-20} \pm 28$ & 25 \\ 
 SELEX~\cite{Iori:2007pw} & $65 \pm 13 \pm 9$ & 83 \\ 
 \hline
\end{tabular}\end{center}

\end{document}